\documentclass[sigconf]{acmart}
\AtBeginDocument{%
  }

\usepackage{tabularx}
\usepackage{algorithmic}
\usepackage{graphicx}
\usepackage{textcomp}
\usepackage{multirow}
\usepackage{rotating}
\usepackage{makecell}
\usepackage{float}
\usepackage{url}
\usepackage{colortbl}
\usepackage{hhline}

\definecolor{Anakiwa}{rgb}{0.576,0.737,1}

\setcopyright{acmlicensed}
\copyrightyear{2026}
\acmYear{2026}
\setcopyright{cc}
\setcctype{by}
\acmConference[WSESE '26]{3rd International Workshop on Methodological Issues with Empirical Studies in Software Engineering }{April 12--18, 2026}{Rio de Janeiro, Brazil}
\acmBooktitle{3rd International Workshop on Methodological Issues with Empirical Studies in Software Engineering (WSESE '26), April 12--18, 2026, Rio de Janeiro, Brazil}
\acmDOI{10.1145/3786149.3788307}
\acmISBN{979-8-4007-2382-7/2026/04}

\begin{document}

\title[Operationalizing Software Engineering Theories for Practical Validation]{Operationalizing Software Engineering Theories for Practical Validation}

\author{Isaque Alves}
\affiliation{%
  \institution{University of São Paulo}
  \city{São Paulo}
  \country{Brazil}
}
\email{isaque.alves@ime.usp.br}

\author{Fabio Kon}
\affiliation{%
  \institution{University of São Paulo}
  \city{São Paulo}
  \country{Brazil}
}
\email{kon@ime.usp.br}

\author{Jessica Diaz}
\affiliation{%
  \institution{Universidad Politécnica de Madrid}
  \city{Madrid}
  \country{Spain}
}
\email{yesica.diaz@upm.es}

\author{Carla Rocha}
\affiliation{%
  \institution{University of Brasília}
  \city{Brasília}
  \country{Brazil}
}
\email{caguiar@unb.br}

\renewcommand{\shortauthors}{Alves et al.}

\begin{abstract}
\textit{Context:} Software Engineering often adapts theory-building frameworks from the social sciences to address socio-technical complexity. The key phases of the theory-building process are conceptual development, operationalization, testing, and application.  Operationalization translates abstract concepts into measurable elements for empirical validation. This phase is essential for delivering the practical utility required by an applied science like Software Engineering. \textit{Objective:} We propose a systematic procedure for the operationalization phase that bridges the gap between abstract concepts and empirical validation, ensuring the resulting theory is both rigorous and practically useful.\textit{Method:} We extend the operationalization framework proposed by Sjøberg et al. and formulate non-causal hypotheses following Dubin’s approach. Our procedure defines variables, selects indicators, and systematically derives hypotheses. \textit{Results:} We present a replicable, evidence-based methodological guideline that preserves a clear chain of evidence and supports practical validation. We illustrate the procedure using the DevOps Team Taxonomies Theory. \textit{Conclusion:} This guideline provides a transparent chain of evidence from theory to testable elements, empowering researchers to ground theoretical advancements in empirical evidence and deliver actionable insights for practitioners.


\end{abstract}





\maketitle

\section{Introduction}

Theory, as a “coherent description, explanation, and representation of observed phenomena”~\cite{gioia1990}, is continuously shaped through the process of producing, confirming, and adapting. Software Engineering (SE), as a socio-technical discipline, requires theory-building methodologies that integrate human, organizational, and technical factors. Unlike social fields with established theory-building foundations~\cite{lynham2002, dubin:1978, charmaz:2014}, SE is relatively new in its systematic methodological guidelines to theory-building, and researchers often adapt methods from the social sciences. Sjøberg et al.~\cite{sjoberg:2008}, for example, expanded and tailored Lynham’s~\cite{lynham2002} theory-building method to address the socio-technical nature of SE. Sjøberg’s framework presents an iterative cycle for theory-building, encompassing: (1) Conceptual Development, which designs theories at a conceptual level, (2) Operationalization, which translates these concepts to the practical domain and enables, (3) Testing, and (4) Application~\cite{sjoberg:2008}.

In applied disciplines such as SE, theories gain greater relevance and impact when they can be effectively implemented and tested by practitioners~\cite{lynham2002}. To accomplish this, the operationalization phase enables researchers to apply and test abstract ideas in real-world scenarios by translating concepts into constructs and propositions into actionable and observable hypotheses~\cite{sjoberg:2008}. For instance, consider the concept of usability in user interfaces. Though inherently abstract, it can be operationalized through measurable indicators such as average task completion time, error rate, and user satisfaction scores obtained via standardized instruments like the System Usability Scale (SUS). These metrics enable empirical evaluation of usability in real-world software systems.

This study contributes to the ongoing research on the continuous development and refinement of software engineering theories. We extend the definitions provided by Lynham and Sjøberg et al. by integrating Dubin’s logic to present a systematic procedure for operationalizing theories. This approach bridges abstract concepts and empirical validation, ensuring that the resulting theory maintains academic rigor while delivering practical utility.

Based on the Sjøberg framework, we outline a structured guideline for operationalizing SE theories, aiming to provide full traceability between the conceptual and practical elements, ensuring a clear and continuous link from abstract concepts to measurable components. We focus on analyzing “what if” scenarios rather than causality or predictive generalization following Dubin’s non-causal modeling method~\cite{dubin:1978} and Pérez et al.~\cite{perez2024theory} in identifying interactions between concepts as the foundation for operationalization. Dubin emphasized that many typically used causal relationships are merely sequential interactions, noting, for example, that an alarm clock ringing before sunrise does not cause the sun to rise~\cite{jaccard2019theory}.

To illustrate and discuss the systematic procedure, we apply the operationalization process to the DevOps Team Taxonomies Theory (T3), developed in a collaboration between teams from the University of São Paulo and Universidad Polit\'ecnica de Madrid. We transform this theory from a conceptual model into clear and precise elements that enable researchers to test, apply, and potentially contradict it with evidence~\cite{popper2005logic}, represented as a set of 83 testable hypotheses to be confirmed or refuted at the practical level. This paper thus advances the understanding and application of the operationalization process in theory-building within SE by detailing its execution. We discuss how the hypotheses and propositions facilitate traceable empirical theory evaluation and provide a foundation for future studies on team dynamics in socio-technical systems within SE. 

\section{Continuous Theory-building}
\label{sec:ctb}

\begin{figure}
  \centering
  \includegraphics[width=1\linewidth]{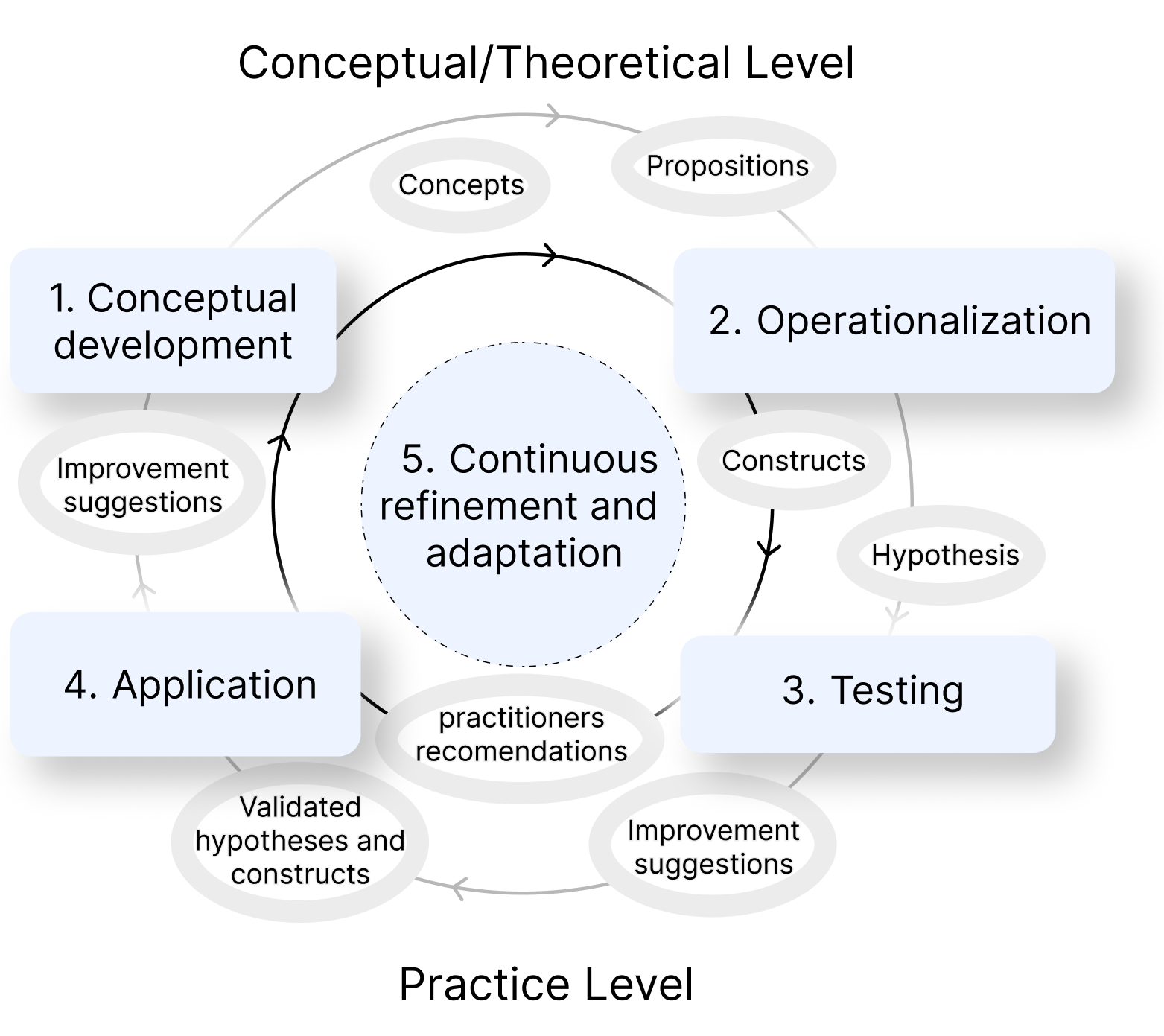}
  \Description{Methodological framework for theory-building}
  \caption{Methodological framework for theory-building.}
  \label{fig:general_process}
\end{figure}

Lynham~\cite{lynham2002} defines theory-building as a structured method that involves both “theorizing to practice” and “practice to theorizing”. The process is “continuous” because findings from empirical testing and application consistently feed back into the conceptual framework, ensuring the theory is constantly refined and adapted to remain relevant over time. In this work, we adopted and refined the procedures proposed by Sjøberg’s methodological framework~\cite{sjoberg:2008}, which defines theory-building as a continuous process of refinement and adaptation, comprising five stages we depict in Figure~\ref{fig:general_process}: 

\begin{itemize} \item \textbf{1. Conceptual Development} generates theory by identifying and defining concepts and their relationships to describe the phenomenon. It relies on inductive and abductive processes to synthesize data and establish concepts and the relationship among them (propositions) to describe or explain a phenomena~\cite{sjoberg:2008}.

\item \textbf{2. Operationalization} translates conceptual theory into observable and measurable components. It transforms abstract concepts into constructs by assigning empirical values and converts theoretical propositions into testable hypotheses, preparing the theory for empirical evaluation.

\item \textbf{3. Testing} involves empirically evaluating the theory by confirming or disconfirming its hypotheses and constructs. Researchers conduct empirical studies to assess the theory’s predictions and refine its elements based on the results~\cite{lynham2002, sjoberg:2008}. We commonly employ surveys and interviews to collect empirical data, confirming or refuting testable hypotheses.

\item \textbf{4. Application} aims to observe the theory in practice. This phase assesses the theory’s relevance and effectiveness in addressing practical problems within specific contexts~\cite{lynham2002}. Researchers commonly use case studies and action research as primary techniques to investigate the application and practical implications of the theory in these contexts.

\item \textbf{5. Continuous Refinement and Adaptation} focus on refining and adapting the theory based on new evidence. Refinement involves clarifying or expanding existing concepts and relationships, while adaptation introduces new elements to ensure the theory evolves alongside advances in knowledge and practice. 

\end{itemize}

Theory Conceptual Development in SE is well-established and can follow a range of methodological guidelines such as: Systematic Literature Reviews (SLR), from secondary studies; Grounded Theory (GT) and case studies, which build theories directly from qualitative data; single-source primary studies, which rely on data from a specific case or organization; and even personal experience, where practitioners distill their knowledge into conceptual frameworks
~\cite{ralph2018toward}. GT and case-study-based approaches have emerged as prominent methods for crafting robust conceptual frameworks grounded in empirical data. GT systematically derives taxonomies and process theories by closely aligning theoretical constructs with real-world observations, thereby minimizing bias~\cite{hoda2022, strauss:1990, charmaz:2014}. Theories can take shape as taxonomies, classifications, process models, or ontologies, each serving distinct analytical and practical purposes. Despite offering solid conceptual foundations, software engineering theories frequently lack systematic procedures for operationalization and measurement in practice.

The problem of lacking operationalization goes far beyond the inability to test and refine the theory. The main consequence is the theory’s incapacity to guide practice and adoption effectively. Without operationalization, theories remain conceptual, failing to provide actionable insights (such as the specific nuances of socio-technical aspects found in DevOps), which leaves adoption vulnerable to simplistic or arbitrary approaches. To address the limitations, we outline a systematic operationalization procedure that can be applied regardless of the theory-building procedures adopted. We extend existing definitions provided by Lynham\cite{lynham2002} and Sjøberg\cite{sjoberg:2008} to present a method for operationalizing theories, thereby ensuring their validity, traceability, and utility in practice.

\section{Theory Operationalization Methodology}
\label{sec:Operationalization}

Theory operationalization addresses two main steps (\cite{sjoberg:2008}, p. 327): (A) operationalizing theoretical concepts into empirical variables and (B) operationalizing theoretical propositions into empirically testable hypotheses.

\begin{figure*}[ht]
  \centering
  \includegraphics[width=\linewidth]{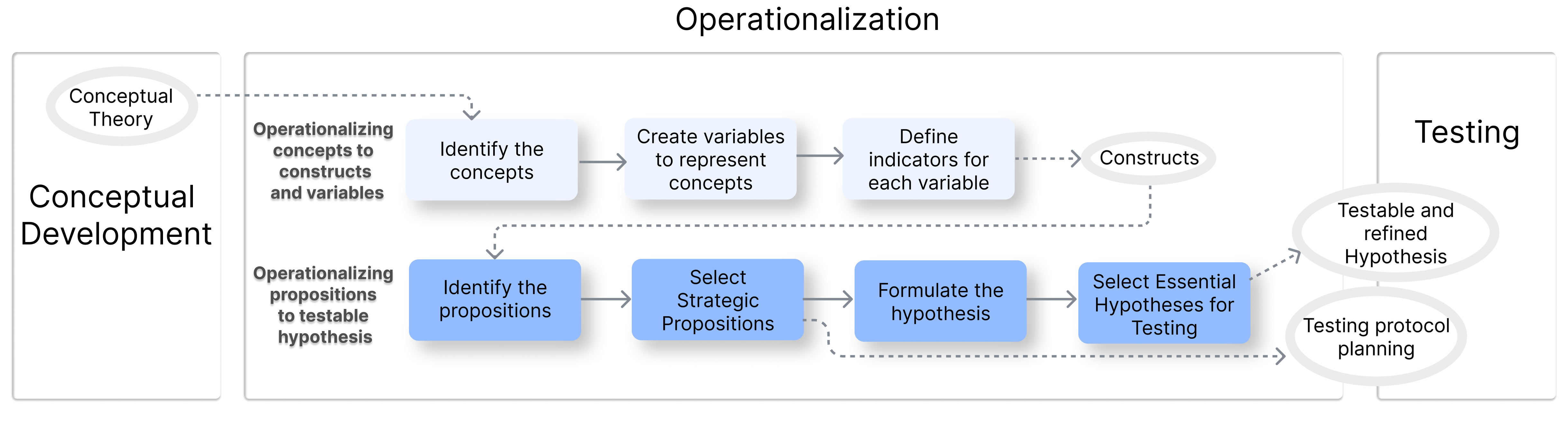}
  \caption{Operationalization requires two steps: (A) defining concept variables and indicators, and (B) translating propositions into strategic, testable hypotheses. This results in refined hypotheses and an empirical testing protocol.}
  \Description{Operationalization requires two steps: (A) defining concept variables and indicators, and (B) translating propositions into strategic, testable hypotheses. This results in refined hypotheses and an empirical testing protocol.}
  \label{fig:detailed_process}
\end{figure*}

Figure~\ref{fig:detailed_process} illustrates the detailed operationalization process we propose in this paper, adapted from Dubin~\cite{dubin:1978}, with steps highlighted in a dark blue background indicating adaptations or methods developed for this study. To ensure clarity in the process, we distinguish key theoretical components: constructs, concepts, empirical indicators, propositions, and hypotheses. When we operationalize a concept, it becomes a construct, as it acquires specific definitions regarding measurement through established variables, metrics, and indicators~\cite{sjoberg2022construct}. A hypothesis represents a prediction about the values of the units within a theory, where empirical indicators are used to measure the specified units in each proposition (the relationship among concepts)~\cite{dubin:1978}. 


\subsection{Operationalizing theoretical concepts into empirical variables}

The first step of theory operationalization aims to identify concepts, empirical values from a conceptual theory through three steps~\cite{pritha2022}: (i) identifying the concepts that describe a phenomenon, (ii) selecting one or more variables to represent each concept, and (iii) choosing indicators for each variable. It is also necessary to define the range of values that each indicator may take over time to ensure reliability and consistency.

The operationalization process for this study is deeply rooted in GT. We perform coding and constant comparison, which produces the conceptual relationships between codes and their properties as they emerge. Consolidating these emerging concepts and establishing the conceptual relationships between them into constructs and propositions. This is a meticulous phase where we perform an analysis of all identified concepts, exploring the connections and interdependencies among them.

To identify the concepts, we examine the data, searching for descriptions of relationships. Concepts are typically expressed in relationships or when describing a phenomenon. The next step is to identify attributes that help explain the concepts. Some of these attributes and indicators are presented implicitly in the data and the provided descriptions. Finally, the indicators serve as a refinement that enables the instantiation of a structure or theory. As we describe different types of structures or frameworks, each with distinct characteristics, these characteristics become indicators of their attributes. These indicators enable the attributes to represent distinct and unique outcomes within the given context.

\subsection{Operationalizing theoretical propositions into empirically testable hypotheses}

The subsequent step consists of translating theoretical propositions into testable hypotheses that integrate the previously defined variables and indicators. According to Dubin~\cite{dubin:1978}, “A hypothesis is the predictions about the values of the units of a theory in which empirical indicators are employed for the named units in each proposition.” After defining the variables and indicators, we identify and select the most strategic propositions, focusing on formulating simple and testable hypotheses (see Figure~\ref{fig:detailed_process}). We now detail each aspect of this operationalization, a structured four-step process:

\subsubsection{Identifying the propositions}

A proposition expresses the relationship between two or more concepts. By translating these concepts into variables and indicators in the previous step, we can derive propositions grounded in both data and theoretical reasoning. This allows us to investigate why particular structures emerge, their consequences, and the trade-offs and challenges they entail. 

Some relationships are explicitly stated in the data, such as the proposition presented by Leite et al.~\cite{leite2022theory} (e.g., “API-Mediated Departments reduce bottlenecks”). Others were implicit, requiring inference based on descriptions of causes and consequences. Finally, to ensure traceability and reliability, we should provide quotations and explanations to justify the propositions, ensuring they remain well-grounded in the data.

\subsubsection{Prioritizing Propositions for Testing}

As the number of propositions increases, the number of potential hypotheses grows combinatorially. “The general rule is that a new hypothesis is established each time a different empirical indicator is employed for any of the units designated in a proposition” ~\cite{dubin:1978}. To avoid an explosion in the number of hypotheses, Dubin~\cite{dubin:1978} recommends selecting strategic propositions  -- that establish boundaries/changes in the described system --  to avoid an overwhelming number of hypotheses and follow the principle of parsimony (not multiplying entities unnecessarily). This step is crucial for the testing process, as the selected propositions serve as the input guiding the formulation of the test protocol. The testing questions will be derived from these propositions, and in the context of surveys, interviews, or questionnaires, they will directly reference them.

The selected propositions describe state changes within the system, excluding those that do not contribute additional information about its current state. In practice, this involves selecting propositions that instantiate specific values of empirical indicators, thereby delineating the boundaries or transitions in the described system. The guiding principle is to ensure that the theoretical framework and team structure explicitly express the relationship between key concepts.

\subsubsection{Formulating hypothesis}

To formulate a simple and testable hypothesis, we focus on three aspects: (i) determining the nature of the semantic relationship between concepts, (ii) checking and simplifying the hypothesis complexity, and (iii) applying abductive reasoning to refine hypotheses.

Our methodology proposes formulating non-causal hypotheses to align with a fundamental principle of scientific inquiry. As articulated by the philosopher David Hume~\cite{hume2016enquiry}, we cannot empirically observe a “necessary connection” that links a cause to its effect. Instead, our experience is limited to observing the “constant conjunction” of events. By framing our hypotheses around non-causal relationships, we provide a simpler description of the phenomena we can actually measure and test. This approach shifts the goal from proving causality to the more practical one of collecting evidence to demonstrate that a conjunction of events is regular and uniform. 

Following Dubin’s framework, we classify these non-causal relationship semantics into Categoric, Sequential, or Determinant interactions \cite{dubin:1978}. A \textit{Categoric Interaction} relies on the presence or absence of concepts without implying change or progression. It corresponds to existence relationships, which establish whether one variable requires the presence or absence of another (e.g., “The presence of responsibility sharing is associated with the presence of collaboration”). A \textit{Sequential Interaction} describes a progression over time, where one concept typically precedes another without implying direct causation. It aligns with gradient relationships, where the presence of one variable gradually modifies another’s value (e.g., “As responsibility sharing increases, collaboration tends to shift from occasional to daily”). Finally, a \textit{Determinant Interaction} defines a structured dependency, where one concept’s value systematically changes alongside another’s. This corresponds to correlation relationships, which describe how two variables change together without establishing causation (e.g., “The frequency of collaboration is proportional to the level of responsibility sharing”).

First, we classify relationship semantics into these non-causal types~\cite{dubin:1978}. Second, we can reduce the number of hypotheses for parsimony by limiting them to two variables instead of introducing unnecessary complexity. For example, a complex hypothesis like \textit{IF (A OR B) THEN C} involves three variables (A, B, and C) but can be split into two simpler hypotheses: \textit{IF (A) THEN C or IF (B) THEN C}. However, formulating a complex hypothesis is necessary if \textit{C} only occurs when both A and B are true simultaneously. In this work, we prioritize simple hypotheses to facilitate future testing, as they require verifying only two variables. In contrast, refuting a complex hypothesis like \textit{IF (A AND B) THEN C} demands proving both \textit{A} and \textit{B} are true while showing \textit{C} does not occur.

Finally, we propose to apply abductive reasoning to refine hypotheses by generating plausible explanations that align with observed phenomena~\cite{walton2014abductive}. This process involves evaluating each hypothesis for its plausibility, internal consistency, and semantic alignment, ultimately enhancing the relevance of the hypotheses in real-world contexts~\cite{magnani2011abduction}. Hypotheses deemed unlikely, dysfunctional, or logically inconsistent with the expected behavior described by the propositions are problematic and must be discussed or removed. 

\subsubsection{Selecting Essential Hypotheses for Testing}

After establishing the hypotheses, we instantiate the constructs to identify the specific hypotheses that can represent and test the main scenarios described by the theory. Therefore, it is necessary to revisit the list of hypotheses and select those that describe propositions and a specific testing scenario. This step aims to refine the hypotheses to the representations of the theory and ensure that it aligns with the practical observations. 

\section{Applying the Methodology: DevOps Team Taxonomies Theory (T3)}

This section presents our illustrative application, the DevOps Team Taxonomies Theory (T3) introduced by Diaz et al.~\cite{diaz2022}, with which we outline an operationalization process. This paper is part of a broader research effort focused on the continuous building and refinement of the T3 theory through conceptual development, operationalization, testing, and refinement. The DevOps team taxonomy classifies team structures, providing a foundation for understanding current practices and planning DevOps adoption in the industry.

\subsection{Introducing T3}
\label{sec:case_study}

T3~\cite{diaz2022} harmonizes existing DevOps taxonomies (secondary data), and was produced by combining Scientific Papers~\cite{leite2021,lopez2021,luz2019,macarthy2020,nybom2016, shahin2017, zhou2022cross} and Grey Literature~\cite{skelton2019team, stateofdevops, sregoogledevops} into a unified theory, undergoing a rigorous process of mitigating confirmation bias~\cite{ralph2018toward}. Diaz et al.~\cite{diaz2022} also proposed a systematic process for theory-building that integrates GT, Inter-Coded Agreement (ICA), and Sjøberg’s approach. T3 consists of a comprehensive theory of DevOps Team Taxonomies and follows a methodological guideline ensuring traceability and reproducibility with available data. More details on T3 are available in Section~\ref{sec:dataavailability}, Appendix A. The theory distinguishes four archetypal team-structure instantiations:

\begin{figure*}
  \centering
  \includegraphics[width=1\linewidth]{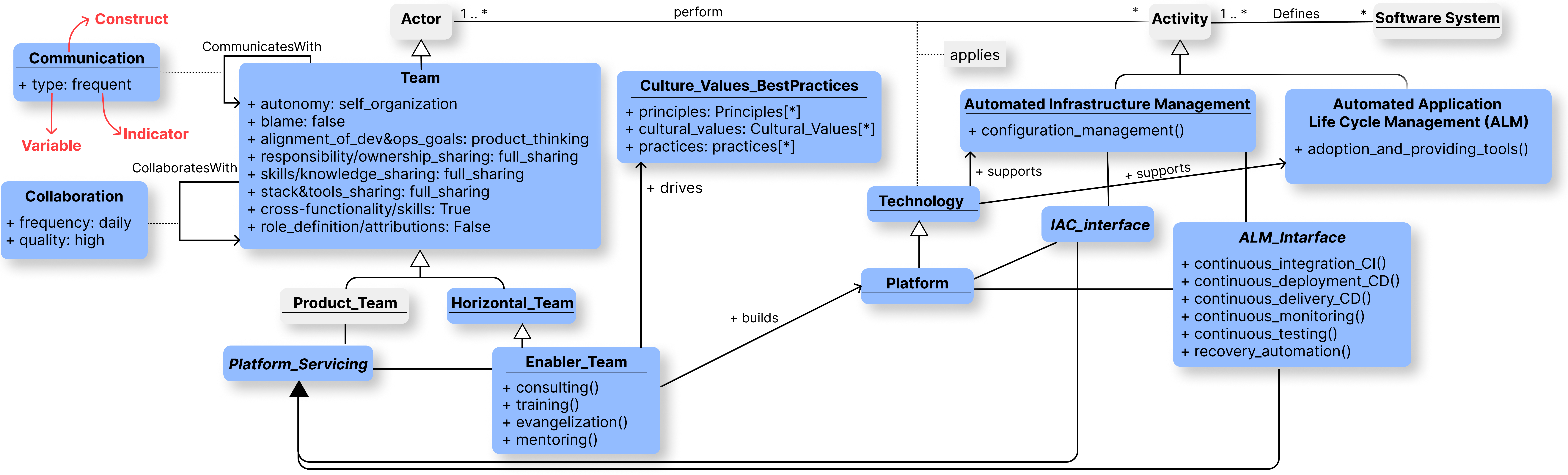}
  \Description{T3 instantiation of an Enabler (Platform) Team.}
  \caption{T3 instantiation of an Enabler (Platform) Team. The team leverages technology to automate workflows and deliver platform services to a Product Team. The communication construct is annotated with captions identifying constructs, variables, and indicators.}
  \label{fig:platform_team}
\end{figure*}

\begin{itemize}
    \item \textbf{Bridge DevOps Team:} helps and supports development and operation teams by deploying and hosting applications in the platforms they build, monitor, and support. The engineers of the Bridge DevOps teams are the DevOps practices facilitators; hence, they create, deploy, and manage the infrastructure (environments) and the deployment (CI/CD) pipelines. They are usually the bridge interface between developers and IT Operations, driving the DevOps values and practices. 
    
    \item \textbf{Enabler DevOps Team:} Organizations create specific teams to satisfy product team necessities. It includes platform servicing and tools (mainly for infrastructure and deployment pipelines), consulting, training, evangelization, mentoring, and human resources. Thus, they behave as enabler teams by providing these capabilities. The Enabling DevOps Teams are named in different ways, e.g., DevOps Centers of Excellence, chapters, guilds, platform/SRE teams, among others.

    \item \textbf{Product Team:} is entirely responsible for and has complete autonomy over a product or service, including scoping, managing, architecting, building, and operating it. This approach allows the product team to innovate quickly with a strong customer focus by aligning development and operations objectives with business goals. 
    
    \item \textbf{Development and Operations Teams:} have well-defined and differentiated roles with their departments and objectives. The development team, for example, focuses on implementing features and is led by a project manager. The interaction between teams typically occurs as a transfer of work. Developers deliver the finished code to the operations team, which creates and manages the infrastructure while also being responsible for the deployment.
\end{itemize}

According to Gregor’s classification~\cite{gregor2006nature},  T3 is an Analysis theory, providing descriptions and conceptualizations of “what is”. Furthermore, due to our epistemological positioning, and since the theory is constructed based on qualitative analysis of a set of data, it must necessarily be limited to a substantive (local) theory, as opposed to a formal (all-inclusive) theory~\cite{glaser:1967}.

\subsection{Operationalizing T3}
\label{sec:operationalizingT3}

In this section, we apply the proposed operationalization methodology outlined in Figure~\ref{fig:general_process} to the DevOps taxonomies in T3. 

\subsubsection{Operationalizing theoretical concepts into constructs}

The operationalization process begins by translating a theory’s concepts into measurable constructs. This phase involves a structured analysis of the raw data and a traceability to the elements of operationalized theory. Here, as an example, we present Team construct, a foundational element of the DevOps theory. 

The \textit{Team} concept is central to the theory; it organizes and defines relationships between different entities. Specializations, such as the \textit{Development Team, Operation Team, and Product Team}, further justify \textit{Team} as an abstract construct representing various organizational structures. Quotations such as “Collaboration and communications among \textit{team} members can considerably increase by establishing cross-functional \textit{teams}” help justify the concept of Team as a constructor. To define the \textit{Autonomy} attribute, we search the data for information that qualifies and describes the constructors. We captured the Autonomy concept through the code “team self-organization \& autonomy”. The indicator values (dependent and self-organization) directly link to specific team types found in the data. A “Product Team” is described as being “cohesive, small and multidisciplinary” with “a high level of sharing of the product ownership,” which is associated with “self-organization”. On the other hand, traditional teams have siloed structures where teams are often “dependent” due to “well-defined and differentiated roles” and have a “transfer of work” rather than shared responsibility.

During the T3’s conceptual development, we apply the framework for describing SE theories proposed by  Sjøberg~\cite{sjoberg:2008}, who already suggests executing Part A of the operationalization (see Figure\ref{fig:detailed_process}) as part of theory building. Therefore,  T3 specifies the DevOps Team taxonomies using 10 constructs and 19 variables, depicted in a UML class diagram. Table~\ref{tab:constructs_variables} lists the team and collaboration constructors followed by their variables and indicators. Additionally, we bring 28 propositions (See Table~\ref{tab:propositions}) that establish relationships between variables and their indicators. All evidence of this process can be found in Section~\ref{sec:dataavailability}.

Finally, Figure~\ref{fig:platform_team} illustrates the \textsf{Enabler (Platform) Team} instantiation of T3. The \textsf{Team} construct is one of the elements explaining the phenomenon of DevOps team structures. This construct includes several associated variables, such as \textsf{autonomy, blame culture}, and \textsf{role definition}. When instantiated, each variable is assigned a specific indicator.
For example, in the case of Enabler (Platform) Teams in Figure~\ref{fig:platform_team}, the Team construct is characterized by \textsf{self-organization autonomy, no blame culture}, and \textsf{full sharing of skills, knowledge, stack, and tools}. These teams also maintain \textsf{high quality} and engage in \textsf{daily} collaboration.

\begin{table}[h]
    \caption{Example of Constructs, variables, and indicators derived from Team and Collaboration concepts (from~\cite{diaz2022}).}
    \Description{A table listing constructs, variables, and indicators. The Team Structure construct includes variables like autonomy and responsibility sharing. The Collaboration construct includes frequency and quality.}
    \label{tab:constructs_variables}
    \begin{tabularx}{\linewidth}{l p{2cm} X}
    \toprule
    \textbf{Constructs} & \textbf{Variables} & \textbf{Indicators} \\
    \midrule
    \multirow{13}{*}{\parbox{2.6cm}{\textbf{Team:} It is an artificial (abstract) concept that represents a team structure of an IT department.}}
      & autonomy & \{dependent, self-organization\} \\
      \cmidrule(l){2-3}
      & blame & \{true, false\} \\
      \cmidrule(l){2-3}
      & alignment of dev \& ops goals & \{local optimization, product thinking\} \\
      \cmidrule(l){2-3}
      & responsibility/ ownership sharing & \{full sharing, medium sharing, minimal or null sharing\} \\
      \cmidrule(l){2-3}
      & skills/knowledge sharing & \{full sharing, medium sharing, minimal or null sharing\} \\
      \cmidrule(l){2-3}
      & stack \& tools sharing & \{full sharing, medium sharing, minimal or null sharing\} \\
      \cmidrule(l){2-3}
      & cross-functionality/ skills & \{true, false\} \\
      \cmidrule(l){2-3}
      & role definition/ attributions & \{true, false\} \\
      \cmidrule(l){2-3}
      & Inherited members & \{product teams, horizontal teams, bridge teams, enabler teams, dev teams, ops teams\} \\
    \midrule
    \multirow{5}{*}{\parbox{2.6cm}{\textbf{Collaboration:} between teams. From the lack or eventual collaboration, to daily collaboration.}}
      & frequency & \{daily, eventual\} \\
      \cmidrule(l){2-3}
      \addlinespace 
      & quality & \{high, low\} \\ \\ \\
    \bottomrule
    \end{tabularx}
\end{table}

\subsubsection{Operationalizing theoretical propositions into empirically testable hypotheses}

The relationships between concepts will shape the hypotheses’ expressions in categoric, sequential, or determinant formats. As presented in Section~\ref{sec:Operationalization}, a new hypothesis is established each time a different empirical indicator is employed. Given a large number of propositions (28), variables (17), and indicators (33), we expect a high number of hypotheses. The indicators developed during the operationalization of theoretical concepts reflect more possible values than actual situations. 

To avoid an explosion in the number of hypotheses, Dubin~\cite{dubin:1978} recommends selecting propositions that contain an instantiation of empirical indicator values establishing boundaries/changes in the described system. In our case, of the 28 propositions, only P26 and P27 are considered non-strategic (see Table~\ref{tab:propositions}). \textsf{Automated Infrastructure Management} and \textsf{Automation Application Life Cycle Management} are specializations of \textsf{Platform Service}. They are taxonomy elements but do not affect the system’s dynamics or the relationships between its concepts.

We applied abductive reasoning to retain only the hypotheses most likely to describe real situations, favoring parsimony. For instance, the absence of \textsf{responsibility/ownership sharing} in a team implies that \textsf{collaboration} may be neither \textsf{daily} nor \textsf{high quality}. This process reduced the initial set of hypotheses from 115 to 83, and further refinement for specific contexts—such as the Enabler (Platform) Team—resulted in 30 hypotheses.

\textit{P1. A team culture based on responsibility/ownership sharing enables collaboration} establishes a categoric relationship between \textsf{team} and \textsf{collaboration}. The first concept, \textsf{team}, has nine associated variables (see Table~\ref{tab:constructs_variables}), and we will focus in: \textsf{responsibility/ownership sharing}. This variable has three associated values: \textsf{full sharing}, \textsf{medium sharing}, and \textsf{minimal or null sharing}. The second concept, \textsf{collaboration}, has two associated variables, each with two values. Therefore, the number of possible hypotheses is 12 (Table~\ref{tab:p1}).

\begin{table}
  \caption{Examples of propositions provided by T3 (from~\cite{diaz2022}).}
  \Description{A table listing propositions from the T3 theory. It shows IDs like P1 and P28 alongside their text descriptions relating concepts like team culture and collaboration.}
  \label{tab:propositions}
  \begin{tabularx}{\linewidth}{l X}
    \toprule
    \textbf{ID} & \textbf{Proposition}\\
        \midrule
        P1 & A \textit{team culture} based on \textit{responsibility/ownership sharing} enables \textit{collaboration}.\\
        ... & ...\\
        P26 & \textit{Automated application life-cycle management} is a \textit{platform servicing}\\
        P27 & \textit{Automated infrastructure management} is a \textit{platform servicing}\\
        P28 & \textit{Enabler (platform) teams} provide \textit{automated application} \textit{life-cycle management}\\
        \bottomrule
    \end{tabularx}
\end{table}

\begin{table}
  \centering
  \caption{Hypotheses Derived from Proposition \textit{P1.} The hypotheses, highlighted in blue, represent the Enabler (Platform) Team structure.}
  \label{tab:p1}
  \small
  \begin{tabular}{|c|c|c|c|c|c|}
    \hline
    \multicolumn{3}{|c|}{\multirow{3}{*}{\textbf{P1 - categoric}}} & \multicolumn{3}{c|}{Team} \\
    \cline{4-6}
    \multicolumn{3}{|c|}{} & \multicolumn{3}{c|}{\begin{tabular}[c]{@{}c@{}}responsibility/ownership\\sharing\end{tabular}} \\ \cline{4-6}
    \multicolumn{3}{|c|}{} & \begin{tabular}[c]{@{}c@{}}full\\ sharing\end{tabular} & \begin{tabular}[c]{@{}c@{}}medium\\ sharing\end{tabular} & \begin{tabular}[c]{@{}c@{}}minimal or\\ null sharing\end{tabular} \\ \hline
    \multirow{4}{*}{\begin{tabular}[c]{@{}c@{}}Colla-\\ boration\end{tabular}} & \multirow{2}{*}{\begin{tabular}[c]{@{}c@{}}fre-\\ quency\end{tabular}} & daily & \cellcolor{Anakiwa} h1.1 & h1.2 & h1.3 \\ \cline{3-6}
     &  & eventual & h1.4 & h1.5 & h1.6 \\ \cline{2-6}
     & \multirow{2}{*}{quality} & high & \cellcolor{Anakiwa} h1.7 & h1.8 & h1.9 \\ \cline{3-6}
     &  & low & h1.10 & h1.11 & h1.12 \\ \hline
  \end{tabular}
\end{table}

Nevertheless, the semantics of the relationship indicate that some form of \textsf{responsibility/ownership sharing} must exist (we discard h1.3, h1.6, h1.9, and h1.12) and that its existence implies an increase in collaboration (gradient relationship). In this case, the concept of \textsf{collaboration} has two variables, each with two possible values. Therefore, we formulate the simple testable hypotheses such as:

\begin{center}
\begin{tabularx}{\linewidth}{|X|}
\hline
\textbf{H1.1 (h1.1 and h1.4):} A team culture based on the full sharing of responsibilities makes it possible to move from eventual collaboration between team members to daily collaboration.\\
\hline
\end{tabularx}
\end{center}

\begin{figure}
  \centering
  \includegraphics[width=1\linewidth]{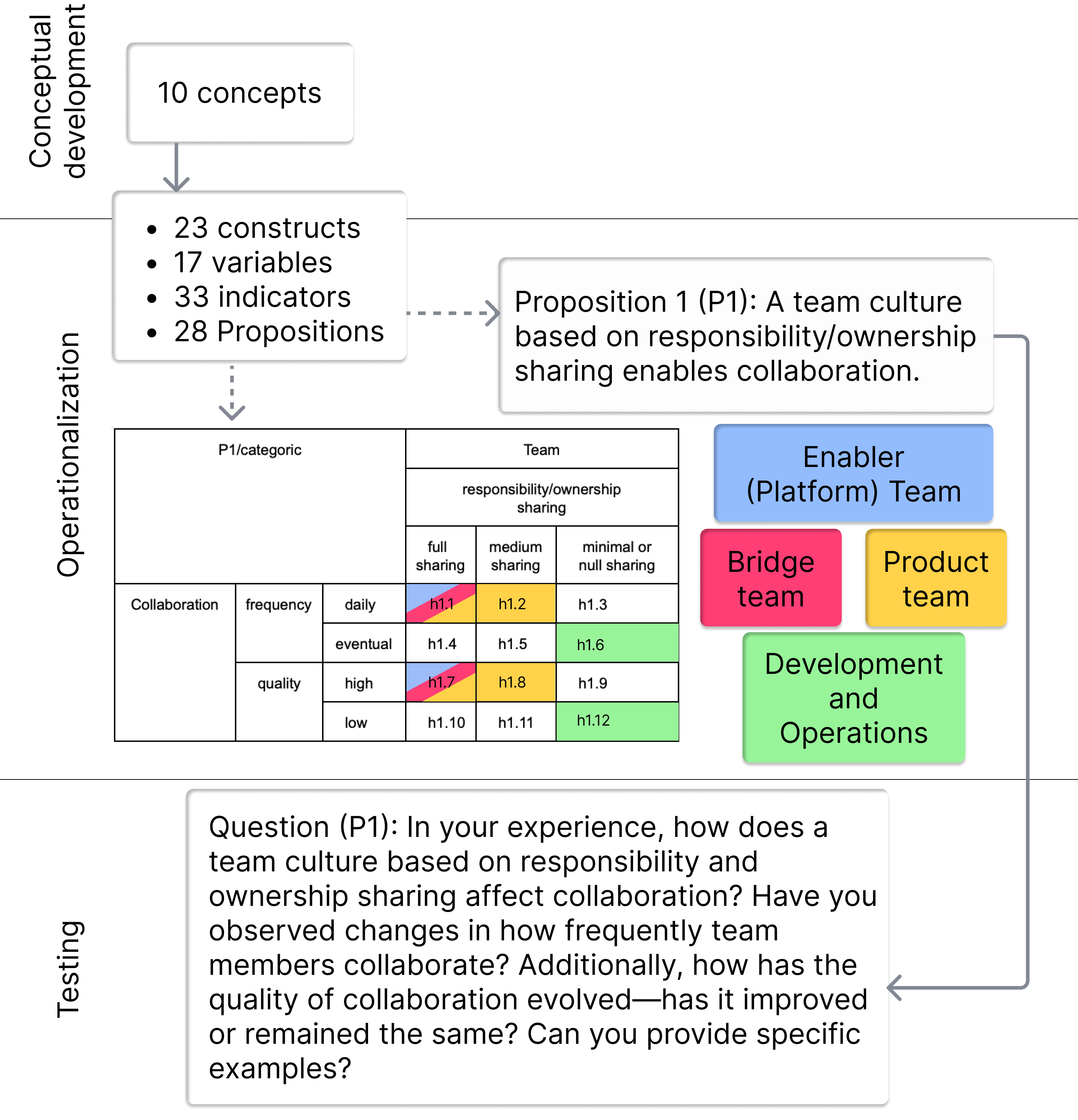}
  \caption{The path from concepts to testable hypotheses (via constructs, variables, indicators, and propositions). P1 is used to generate the research questions, and the hypothesis table is color-coded by team structure.}
  \Description{The path from concepts to testable hypotheses (via constructs, variables, indicators, and propositions). P1 is used to generate the research questions, and the hypothesis table is color-coded by team structure.}
  \label{fig:operationalization_results}
\end{figure}

Figure~\ref{fig:operationalization_results} summarizes the results of T3’s operationalization and illustrates how propositions and hypotheses are applied in the testing phase, making the link between operationalization and testing explicit. In this example, we test the theory with semi-structured interviews: propositions guide the interview protocol, expert responses are analyzed to confirm or refute the hypotheses, and full traceability identifies which theory elements require revision when a hypothesis is refuted.

\section{Discussions and Implications}
\label{sec:dis_implic}

In this paper, we operationalize T3, enabling its continuous testing and refinement. The supplementary material provides the instantiation of four DevOps team structures, detailing their key concepts, variables, and indicators. In this section, we discuss the results, the role of operationalization, and outline the key implications of this work.

\subsection{Practical Utility}

Many organizations aspire to have high-performance DevOps teams; however, it is often difficult to accurately assess their current state or determine how to progress toward higher performance. The DevOps Research and Assessment (DORA) metrics provide a standardized framework to evaluate the effectiveness of DevOps practices, focusing on four key indicators: Lead Time for Changes, Deployment Frequency, Change Failure Rate, and Time to Restore Service (MTTR)~\cite{forsgren2018accelerate}. While these metrics effectively capture observed performance and maturity levels, they fall short of offering actionable guidance.

Consider an organization with an Enabler (Platform) Team that supports product teams yet struggles to increase deployment frequency -- a DORA throughput metric typically associated with shorter lead time for changes. DORA metrics show progress, but not how to achieve them or which practices to adopt. Through operationalization, this issue maps to the Collaboration and Automation constructs. One scenario analysis could reveal limited responsibility and ownership sharing across teams, hindering automation. Consequently, the organization can implement targeted Enabler-Team interventions -- such as culture dissemination, consulting, and seamless tool integration -- to strengthen Shared Responsibility. Greater shared ownership fosters more frequent, higher-quality collaboration, enabling deeper automation, higher deployment frequency, and ultimately, reduced lead time for changes.

In this context, operationalization enables practitioners to (1) assess organizational maturity through defined attributes and variables, effectively conducting an organizational diagnosis; (2) outline a clear path for adoption or evolution by identifying which practices require change (e.g., increasing deployment frequency, reducing silos); and (3) establish indicators and metrics to monitor the effectiveness of these changes.

\begin{center}
\begin{tabularx}{\linewidth}{|X|}
\hline
\textbf{Implication \#1} Operationalization defines clear, measurable constructs that make a theory’s assessment criteria explicit and translate them into concrete, step-by-step guidelines for improving performance and advancing organizational maturity.\\
\hline
\end{tabularx}
\end{center}

\subsection{Traceability and Continuous Adaptation}

Operationalization formally connects the conceptual to the practical and testable levels, defining constructs, variables, and hypotheses (see Figure \ref{fig:operationalization_results}), ensuring empirical traceability. This traceability underpins academic rigor by creating a clear chain of evidence and supporting robust, replicable results. It also enables researchers to adjust theoretical constructs based on empirical findings accurately. For example, when a hypothesis requires revision, traceability clarifies where and how these adjustments should be made.

Subsequent movements -- such as DevSecOps, MLOps, and AIOps -- extend the DevOps paradigm by introducing additional quality dimensions and team dynamics into the delivery lifecycle, including security, data governance, and intelligent automation. From a theory-building perspective, these movements should not be regarded as isolated frameworks, but rather as conceptual evolutions within the DevOps taxonomy, inheriting its foundational constructs of team autonomy, continuous delivery, and cross-functional collaboration. Through operationalization, it becomes possible to derive new theoretical instantiations (e.g., MLOps or DevSecOps) as contextual refinements or “forks” of the core DevOps theory. This approach preserves theoretical continuity while enabling the systematic adaptation of constructs, variables, and indicators to domain-specific concerns, thereby ensuring cumulative knowledge development and empirical testability.

\begin{center}
\begin{tabularx}{\linewidth}{|X|}
\hline
\textbf{Implication \#2:} Viewing DevSecOps, MLOps, and AIOps as evolutions of the DevOps taxonomy -- rather than separate frameworks -- supports cumulative theory-building in socio-technical systems. This approach allows researchers to extend existing constructs and indicators through systematic operationalization, maintaining theoretical continuity while adapting to new domains.\\
\hline
\end{tabularx}
\end{center}

\subsection{Operationalization for the Testing Protocol}

The value of an abstract taxonomy versus an operational one becomes clear when designing a testing protocol. For example, Leite et al.\cite{leite2021organization} characterize siloed departments (i.e., Development and Operations Teams) through properties such as “Developers and operators have well-defined and differentiated roles.” While this descriptive taxonomy offers strong conceptual grounding, its abstract nature complicates the design of measurable tests. In a subsequent study, Leite et al.~\cite{leite2022theory} employed open-ended interviews to validate their model, which increases the analytical rigor required to interpret unstructured data and relate it to abstract theoretical constructs.

On the other hand, operationalization guides the testing protocol, enabling questions based on the propositions and validations through a non-causal hypothesis. For example, we can validate and discuss aspects of teams and collaboration (see Figure \ref{fig:operationalization_results}) without requiring a contextualization and explanation of the entire theoretical framework, thereby avoiding biases in data collection through surveys or interviews. The non-causal hypotheses enhance methodological rigor by moving the goal from proving absolute causation (e.g., Daily team collaboration reduces organizational silos) to the description of events (e.g., Teams with daily collaboration are associated with fewer organizational silos). However, daily collaboration is not the only factor responsible for reducing silos; other aspects, such as strong leadership support, standardized communication protocols, or shared performance metrics, may also contribute significantly to reducing silos. 

\begin{center}
\begin{tabularx}{\linewidth}{|X|}
\hline
\textbf{Implication \#3:} Operationalization serves as a methodological guide for the testing phase. Defining measurable constructs and non-causal hypotheses, it provides structure and direction to testing procedures, ensuring a transparent linkage between empirical findings and theoretical propositions. This systematic connection enables researchers to identify which aspects of a theory are empirically supported and which require revision, thereby strengthening the iterative process of continuous theory-building.\\
\hline
\end{tabularx}
\end{center}

\section{Data Availability}
\label{sec:dataavailability}
All data related to this research providing a detailed chain of evidence, is available at \textbf{\url{https://bit.ly/Operationalization}}.

\section*{Acknowledgment}

This study was financed in part by the Coordenação de Aperfeiçoamento de Pessoal de Nível Superior - Brasil (CAPES) - Finance Code 001, CNPq proc. 308327/2021-7, and FAPESP proc. 2023/00811-0.

\bibliographystyle{ACM-Reference-Format}
\bibliography{bibliography}

\end{document}